%% file: synch_1.tex
\documentclass[conference,comsoc]{IEEEtran}

\usepackage{tikz}
\usepackage{float}
\usepackage{amsthm}
\usepackage{amsmath}
\usepackage{amssymb}
\usepackage{commath}
\usepackage{datetime}
\usepackage{algorithm}
\usepackage{algorithmic}
\usepackage{tikz}
\usetikzlibrary{bayesnet}
\usepackage{multicol}
\usepackage[noadjust]{cite}
\usepackage{ragged2e}
\usepackage{hyperref}
\hypersetup{
	pdfstartview=FitH,
	colorlinks=false,
	linkbordercolor=black,
	pdfborderstyle={/S/U/W 0}
}
\usetikzlibrary{patterns,positioning,fit,shapes,arrows}

\input{Macros}

\IEEEoverridecommandlockouts
\begin{document}
\title{Synchronization in 5G: a Bayesian Approach
\thanks{The research leading to these results has received funding from the European Union's Framework Programme Horizon 2020 for research, technological development and demonstration under grant agreement No. 762057 (5G-PICTURE).}
}

\author{\IEEEauthorblockN{Meysam Goodarzi\rlap{\textsuperscript{\IEEEauthorrefmark{1}}},\,\textsuperscript{\IEEEauthorrefmark{2}} Darko Cvetkovski\rlap{\textsuperscript{\IEEEauthorrefmark{1}}},\,\textsuperscript{\IEEEauthorrefmark{2}} Nebojsa Maletic\rlap{\textsuperscript{\IEEEauthorrefmark{1}}},\, Jes{\'u}s Guti{\'e}rrez\rlap{\textsuperscript{\IEEEauthorrefmark{1}}}, and Eckhard Grass\rlap{\textsuperscript{\IEEEauthorrefmark{1}}},\,\textsuperscript{\IEEEauthorrefmark{2}}}
\IEEEauthorblockA{\IEEEauthorrefmark{1}IHP -- Leibniz-Institut f\"{u}r innovative Mikroelektronik, Frankfurt (Oder), Germany}
\IEEEauthorblockA{\IEEEauthorrefmark{2}Humboldt University of Berlin, Berlin, Germany.}
Emails:\{goodarzi, cvetkovski, maletic, teran, grass\}$@$ ihp-microelectronics.com}
\maketitle
\begin{abstract}
In this work, we propose a hybrid approach to synchronize large scale networks. In particular, we draw on Kalman Filtering (KF) along with time-stamps generated by the Precision Time Protocol (PTP) for pairwise node synchronization. Furthermore, we investigate the merit of Factor Graphs (FGs) along with Belief Propagation (BP) algorithm in achieving high precision end-to-end network synchronization. Finally, we present the idea of dividing the large-scale network into local synchronization domains, for each of which a suitable sync algorithm is utilized. The simulation results indicate that, despite the simplifications in the hybrid approach, the error in the offset estimation remains below 5 ns.

\end{abstract}
%
\IEEEpeerreviewmaketitle
\section{Introduction}\label{sec:intro}
The fifth generation of wireless networks (5G) are expected to deliver a wide variety of services, many of which require the nodes' clocks to be aligned. Distributed beamforming \cite{jagannathan2004effect}, tracking and mobility prediction \cite{wu2010clock,goodarzi2019next}, and localization \cite{zheng2009joint} can be referred to as cases where time synchronization is necessary to guarantee certain levels of quality of service. There has been a great effort to design algorithms for achieving fast and continuous synchronization \cite{levesque2016survey}. Generally, state-of-the-art sync algorithms can be classified into two categories: a) pairwise synchronization \cite{giorgi2011performance, leng2011low, lv2014simulation}, where protocols are primarily designed to synchronize two nodes, and b) network-wide synchronization, where protocols are designed to synchronize a large number of nodes in the network \cite{leng2011distributed, zou2015network, rhee2009clock}. 

 Among all the existing synchronizers, perhaps IEEE 1588 \cite{eidson2002ieee}, often denoted as Precision Time Protocol (PTP), is the most well-known sync protocol employed in a wide variety of applications. PTP along with the Best Master Clock Algorithm (BMCA)
uses hardware time-stamping and pairwise communication between nodes to determine the Master Node (MN) and to perform synchronization. Although this combination might offer satisfactory performance in networks with medium time precision sensitivity, errors in time-stamping on one hand \cite{giorgi2011performance}, and the BMCA failure in determining the MN on the other hand \cite{gaderer2008master}, can lead to a significant deterioration of the performance in time precision sensitive networks. The former results from the layer where the time-stamps are taken, while the latter can be potentially due to the fact that the communication network might be based on a mesh topology. In \cite{giorgi2011performance} and \cite{lv2014simulation} the substantial benefit of Kalman Filtering (KF) has been revealed, whereby the negative impact of the time-stamping error on the sync processes is alleviated. Furthermore, \cite{leng2011distributed} proposes a network-wide synchronization where Factor Graphs (FGs) are used along with Belief Propagation (BP) for the nodes to perform synchronization. Unlike BMCA, in BP the nodes exchange their opinion about each other, thereby reaching an agreement about their clock status even if the network (or its corresponding FG) contains loops.   

While the above-mentioned works have made valuable contributions towards synchronization, it is highly unlikely that each individual solution, e.g. the ones in \cite{giorgi2011performance} and \cite{leng2011distributed}, can alone achieve the high precision aimed by 5G while keeping the complexity low. To bring both types of algorithms together, the idea of a \textit{synchronization harmonizer} has been introduced in \cite{ruffini2017novel}. One of the key aspects thereof is to equip the network with different sync algorithms (or a combination thereof). In fact, in order for the harmonizer to meet the desirable sync precision, one can divide the large scale network into multiple local synchronization domains and employ the suitable sync algorithm based on each local network topology. 
In this manner, it is easier to satisfy the requirement of the relative time error in the sync domains, i.e. each can run the best suitable algorithm based on its topology and capabilities \cite{li2017analysis}. 

The contribution of this paper is summarized as follows:
\begin{itemize}
\item We analyze the statistical relation between the neighboring nodes with the aid of PTP time-stamp exchange.
\item We discuss pairwise and network-wide statistical synchronization algorithms based on KF and FG, respectively.
\item We propose a hybrid approach to achieve high precision time synchronization across the network.
\end{itemize} 
The rest of this paper is structured as follows: In Section II, we introduce our system model and obtain the statistics between the nodes. In Section III, the clock offset estimation based on the obtained statistics is discussed. Furthermore, simulation results are
presented and discussed in Section IV. Finally, Section V concludes this work and indicates the future work.
\subsubsection*{Notation} The boldface capital $\boldsymbol{A}$ and lower case $\boldsymbol{a}$ letters denote matrices and vectors, respectively. $\boldsymbol{1}_N$ is a vector with $N$ entries each equal to $1$. $\boldsymbol{I}_N$ is a $N\times N$ dimensional matrix with all its diagonal elements equal to $1$. The symbol $\propto$ represents
the linear scalar relationship between two real valued functions. $\mathrm{Var}(\cdot)$ and $\mathcal{E}\{\cdot\}$ denote the variance and statistical expectation, respectively. $\mathcal{N}(\mu, \sigma^2)$ represents a Gaussian distribution with mean $\mu$ and variance $\sigma^2$.
\section{System Model}
\subsection{Clock Model}
Each node $i$ is considered to have the clock model 
\begin{equation}
c_i(t) = \gamma_i t + \theta_i
\end{equation}
where $\gamma_i$ and $\theta_i$ denote the clock skew and offset, respectively. Furthermore, $t$ represents the reference time. In fact, function $c_i(t)$ determines how the reference time and clock of node $i$ are mapped onto each other. Given that, the goal of synchronization \footnote{In this work, we focus only on offset estimation and leave the skew estimation for the future works. In fact, the goal of this work is only to reveal the potential performance of hybrid synchronization.
} is to find the 
offset $\theta_i$ (or a transformation thereof) for each node and apply corrections such that, ideally, all the clocks show the same time as the reference time. 
\subsection{Offset Decomposition}
\begin{figure}[t!]
\includegraphics[height=2.3cm,width=0.9\linewidth]{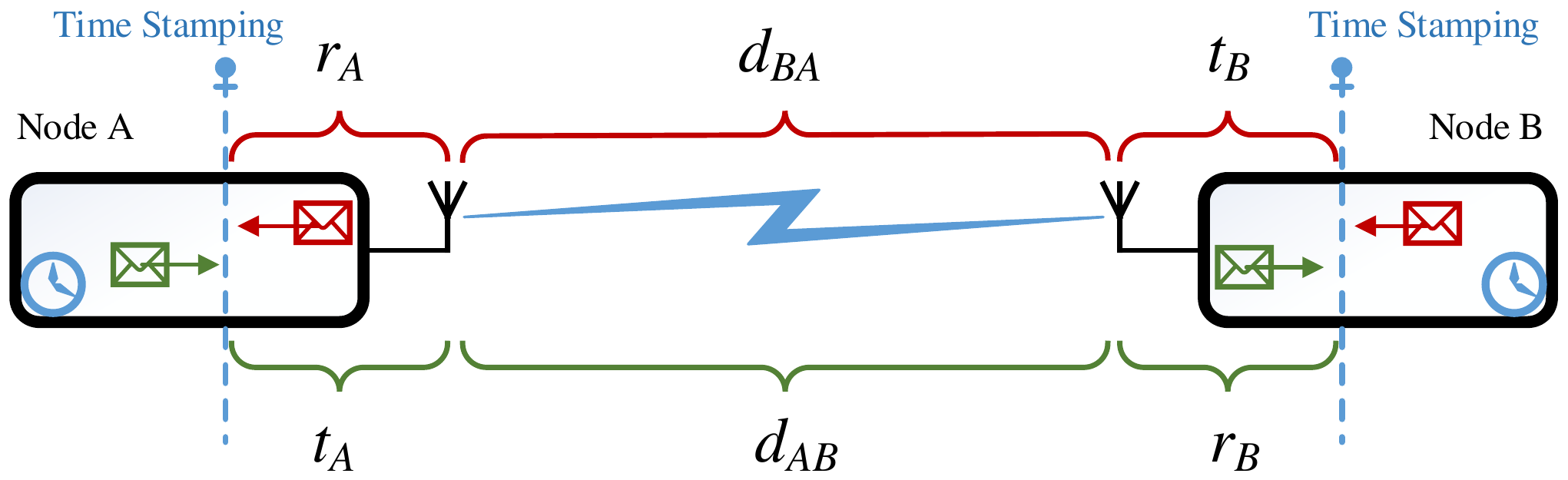}
\centering
\caption{Delay Decomposition.}
\label{fig:deldec}
\end{figure}

To achieve the above-mentioned goal, we begin with decomposing the offset between two nodes, thereby acquiring a reasonable conception of the offset components to be compensated. The offset $\theta_i$ is comprised of several components, as shown in Figure \ref{fig:deldec}. $t_A$ (and $t_B$) is the time that a packet needs to leave the transmitter after being time-stamped \footnote{From now on, the term ``time-stamp" refers to hardware time-stamping.}, $d_{AB}$ and $d_{BA}$ denote the propagation delay, and $r_B$ (and $r_A$) is the time that a packet needs to reach the time-stamping point after arrival at the receiver. In general, the packets sent from node A to node B do not face the same delay as the packets sent from node B to node A. In other words $$t_A + d_{AB} + r_B \neq t_B + d_{BA} + r_A.$$ Furthermore, we can define $T = t_A + r_B,$ and $R = t_B + r_A$.
Generally, $T$ and $R$ are random variables due to several independent random processes and therefore can be assumed i.i.d. Gaussian random variables, whereas $d_{AB}$ and $d_{BA}$ are due to propagation and usually assumed to be deterministic and symmetric ($d_{AB}=d_{BA}$)\cite{leng2011distributed}.
\subsection{Measurement Model}
\begin{figure}[t!]
\includegraphics[width=0.45\linewidth]{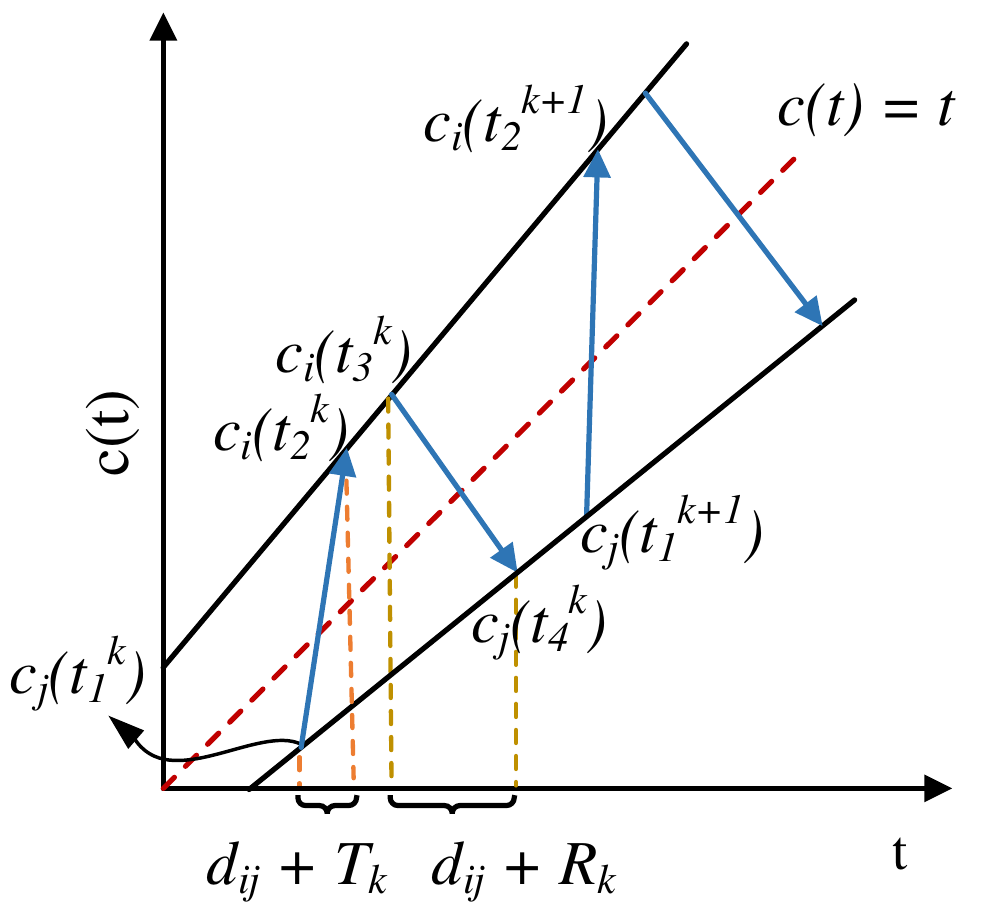}
\centering
\caption{Message exchange between adjacent nodes.}
\label{fig:stamp}
\end{figure}

We use the time-stamping shown in Figure \ref{fig:stamp}, implemented by the PTP protocol \cite{eidson2002ieee}, to estimate the offset between two adjacent nodes. Thus for the $k$-th round of message exchange we can write
\begin{align}
&\alij (c_i(t_{2}^k)-\theta_i) = c_j(t_1^k) - \theta_j + \gamma_j(d_{ij} + T_k)  ,\label{eq:beforeapp1}\\
&\alij (c_i(t_{3}^k)-\theta_i) = c_j(t_4^k) - \theta_j - \gamma_j(d_{ij} + R_k),
\label{eq:beforeapp2}
\end{align}
where $\alpha_{ij} = \frac{\gamma_j}{\gamma_i}$ represents the relative clock skew. 
Generally, the skew of a properly working clock is considered to be close to $1$ \cite{etzlinger2014cooperative}. In fact, given that the term $d_{ij}+ T_k$ and $d_{ij} + R_k$ are expected to be of low value, we can adopt the approximation in \cite{wu2010clock} given by
\begin{align}
\gamma_j(d_{ij} + T_k) \approx (d_{ij} + T_k),\\
\gamma_j(d_{ij} + R_k) \approx (d_{ij} + R_k).
\end{align}  
Consequently (\ref{eq:beforeapp1}) and (\ref{eq:beforeapp2}) turn into
\begin{align}
&\alij  c_i(t_{2}^k) = c_j(t_1^k)  - \theta_j + \alij\theta_i + d_{ij} + T_k  ,\label{eq:tstamp1}\\
&\alij  c_i(t_{3}^k) = c_j(t_4^k)  - \theta_j  + \alij\theta_i - d_{ij} - R_k. \label{eq:tstamp2}
\end{align}
Nevertheless $\alpha_{ij}=\frac{\gamma_j}{\gamma_i}\approx 1$ does not hold since the value of time-stamps $c_i(t_{2}^k)$ and $c_i(t_{3}^k)$ can be very large and therefore even the small amount of $\alpha_{ij}$ could lead to a considerable difference in their multiplication, and consequently in the estimation of the clock offsets. For the sake of simplicity we change the notation of the time-stamps, e.g., $c_j(t_{1}^k)$ is denoted by $\ci{1}{k}.$ Summing up (\ref{eq:tstamp1}) and (\ref{eq:tstamp2}), we can write
\begin{align}
&\alij (\cj{2}{k} + \cj{3}{k}) = \ci{1}{k} + \ci{4}{k} - 2(\theta_j - \alij\theta_i)+ Z_k,
\label{eq:finstmp}
\end{align}
where $\{Z_k = T_k - R_k\}_{k=1}^{k=K}$ are i.i.d. Gaussian random variables with mean $\mu = \mu_{T} - \mu_{R}$ and the variance $\sigma^2 = \sigma_T^2 + \sigma_R^2$. The mean, $\mu,$ is expected to be close to $0$ and therefore negligible \cite{leng2011distributed}. The parameter $\sigma^2$ is assumed to be static as it is mostly due to the hardware properties of the nodes \cite{etzlinger2014cooperative} and can be computed by the nodes in a defined training phase. In the following, we utilize the properties of PTP time-stamping  to estimate the parameters $\alpha_{ij}$ and $\sigma^2.$    
\subsubsection{Relative clock skew $\alij$}
Based on the Simple Skew Clock model introduced in \cite{pasztor2002pc} and employed in \cite{giorgi2011performance, lv2014simulation}, the individual skews, and consequently the corresponding relative skews are assumed to be constant within each sync period.
A simple PTP-based estimation of $\alpha_{ij}$ for a single sync period can be given by
\begin{equation}
\alpha_{ij} = 1 + \mathcal{E}\{\zeta\} ,
\label{eq:drift}
\end{equation}
where 
\begin{equation}
\zeta = \frac{\left(\cj{2}{k}-\cj{2}{k-1}\right)-\left(\ci{1}{k}-\ci{1}{k-1}\right)}{\ci{1}{k}-\ci{1}{k-1}}.
\label{eq:zeta}
\end{equation}
The parameter $\zeta$ denotes the clock drift between the two nodes. In fact, (\ref{eq:drift}) states that the relative skew is equal to the mean of the relative skews in one set of PTP time-stamp exchange (a single set can include $K$ rounds of time-stamp exchange, Figure \ref{fig:stamp}).
\subsubsection{Variance $\sigma^2$} \label{sssec:var} Subtracting (\ref{eq:tstamp2}) from (\ref{eq:tstamp1}) gives
\begin{equation}
 \alij(\cj{2}{k} - \cj{3}{k}) -  (\ci{1}{k} - \ci{4}{k}) = 2d_{ij} + T_k + R_k .
\label{eq:sigma}
\end{equation}
$\sigma^2$ is calculated in the first communication of two nodes in the course of a training phase. In particular, repeating the message exchange during this phase provides the nodes with sufficient samples to calculate the variance of $2d_{ij} + T_k + R_k$ which is $\sigma^2 = \sigma_T^2 + \sigma_R^2$. It is clear that  $\mathrm{Var} (T_k+R_k) = \mathrm{Var} (T_k-R_k)$ since $T_k$ and $R_k$ are independent. 
\subsection{Pairwise Conditional Probability}
The aim here is to define the conditional probability of delay between two adjacent clocks given their offsets, $\theta_i$ and $\theta_j$. Given that $Z_k$ is Gaussian distributed, (\ref{eq:finstmp}) states that the relation between the set of time-stamps, $\cij$ and $\cji,$ and the offset parameters, $\theta_i$ and $\theta_j$, is as follows: 
\begin{multline}
P(\Cij|\theta_i,\theta_j)= \\
  \left(\frac{1}{\sqrt{2\pi}\alij\sigma}\right)^K\exp\left(-\frac{\parallel\alij\cji - \cij - 2(\theta_j-\alij\theta_i)
\cdot\mathbf{1}_K \parallel^2}{2\alij^2\sigma^2}\right)\\ = \left(\frac{1}{\sqrt{2\pi}\alij\sigma}\right)^K \exp\left(-\frac{4K}{2\alij^2\sigma^2}\left[(\alij\theta_i-\theta_j)+ \frac{1}{2K}\mathbf{1}_K^T\Cij\right]^2\right),
\label{eq:condprob}
\end{multline}
where $$\cji = \left[\cj{2}{1}+\cj{3}{1} ,\cdots, \cj{2}{K}+\cj{3}{K}\right],$$ $$\cij = \left[\ci{1}{1}+\cj{4}{1} ,\cdots, \cj{1}{K}+\cj{4}{K}\right],$$ and $$ \Cij = \left[\Cijk{1},\cdots,\Cijk{K}\right] = \alij\cji - \cij.$$
With above conditional probability distribution defined, the Bayesian posterior distribution of clock offset, $\theta_i,$ is given by
\begin{multline}
p(\theta_i|\Cij) = \int p(\theta_i, \theta_j|\Cij) d\theta_j\\
\propto \int p(\Cij|\theta_i, \theta_j)p(\theta_i)p(\theta_j)d\theta_j.
\label{eq:estpair}
\end{multline}
where $p(\theta_i)$ and $p(\theta_j)$ denote the prior distribution of $\theta_i$ and $\theta_j$, respectively, and assumed to be Gaussian \cite{wu2010clock}. Consequently, the clock offset for node $i$ can be estimated as
\begin{equation}
\hat{\theta}_i = \argmax_{\theta_i} p(\theta_i|\Cij) = \argmax_{\theta_i} \int p(\theta_i, \theta_j|\Cij) d\theta_j.
\label{eq:est}
\end{equation}
In the next section, we firstly estimate the clock offsets for the pairwise synchronization. Later on, we extend (\ref{eq:estpair}) and (\ref{eq:est}) for the network-wide synchronization where the offsets are estimated considering the impact of all nodes on each other.
\section{Clock Offset Estimation}
\subsection{Pairwise synchronization based on KF}\label{sec:kf}
\begin{figure}
\includegraphics[width=0.75\linewidth]{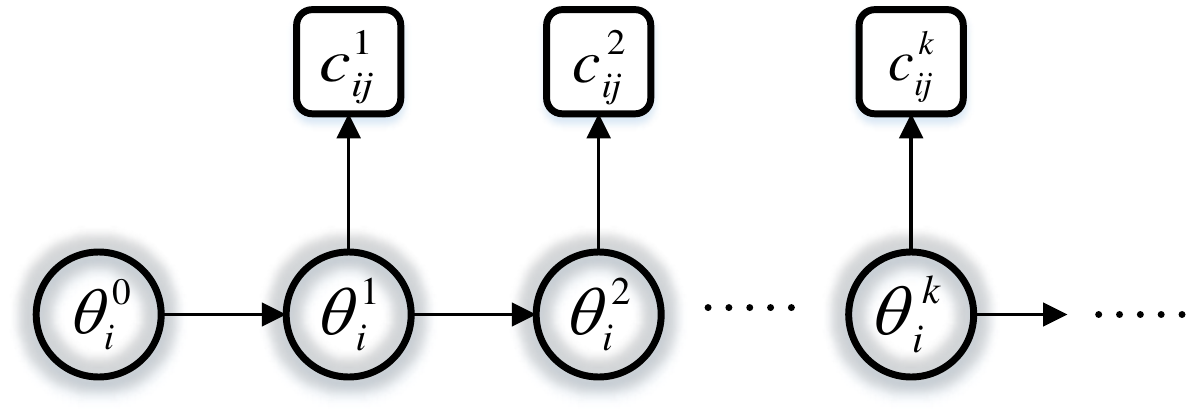}
\centering
\caption{Bayesian representation.}
\label{fig:bayesrep}
\end{figure}

Let us assume that the estimation of $\theta_i$ after $k$-th round of time-stamp exchange is denoted by $\tta{k}$ (Figure \ref{fig:bayesrep}).
The estimation can then be written as
\begin{equation}
p(\tta{k}|\Cij) = \int \jthnc{0}{k}\ d\Theta^{k-1},
\end{equation}
where $\Theta^{k-1} = \left[\tta{0},\cdots,\tta{k-1}\right]$. Employing Bayes rule: 
\begin{equation}
p(\tta{k}|\Cij) \propto \int \jcthn{0}{k}\ttan{0}{k}\ d\Theta^{k-1}.
\label{eq:bayesrule}
\end{equation}
Assuming the Markov property, the terms in the integral can be rewritten as 
\begin{align}
\begin{split}
&\jcthn{0}{k} = \cntn{1}{1}\cdots \cntn{k}{k}, \\
&\ttan{0}{k} = \tnt{k}{k-1}\cdots \tnt{2}{1}p(\theta_i^0).
\end{split}
\label{eq:markov}
\end{align}
Plugging (\ref{eq:markov}) into (\ref{eq:bayesrule}) leads to 
\begin{multline}
p(\tta{k}|\Cij) \propto \\ \underbrace{\int p(\theta_i^0)\left[\prod_{r=1}^{k-1}p(\tta{r}|\tta{r-1})\cntn{r}{r}\right]\tnt{k}{k-1}d\Theta^{k-1}}_{p(\tta{k}|c_{ij}^{0:k-1})}\cntn{k}{k},
\label{eq:longeq}
\end{multline}
which can be simplified as follows:
\begin{equation}
p(\tta{k}|\Cij) \propto p(\theta_i^{k}|\Cijk{0:k-1})\cntn{k}{k}.
\label{eq:bayesfin}
\end{equation}
Assuming Gaussian distribution for the conditional probabilities in (\ref{eq:bayesfin}), the KF equations can readily be derived \cite{barker1995bayesian}. 

In practice, there is always uncertainty in time-stamping, i.e. the time a packet is stamped is different from the time it actually leaves/enters a transmitter/receiver. Given that, it appears necessary to preprocess the measured quantities (offset and drift) in order to obtain their true values. While the estimation in equation (\ref{eq:bayesfin}) can be directly used to preprocess the measured offset and drift \cite{rhee2009clock}, for the sake of clarity and tractability, we present the  KF equations corresponding to (\ref{eq:bayesfin}) which can then be employed to make as precise an estimation as possible \cite{giorgi2011performance}.

We begin with the first KF equation, typically known as prediction equation. It is given by \cite{lv2014simulation}
\begin{align}
\mathbf{x}_{k} = \mathbf{A}\mathbf{x}_{k-1} + \boldsymbol{\omega}_k.
\label{eq:kfpred}
\end{align}
Bringing (\ref{eq:kfpred}) in the context of pairwise synchronization and assuming the process noise is negligible ($\boldsymbol{\omega}_k = 0$)
\begin{align}
\begin{bmatrix}
\theta_i^{k} \\ \zeta_i^{k}
\end{bmatrix} = \begin{bmatrix}
1 & \Delta T \\ 0 & 1
\end{bmatrix} \begin{bmatrix}
\theta_i^{k-1} \\ \zeta_i^{k-1}
\end{bmatrix},
\end{align}
where $\Delta T$ denotes the time needed for one round of time-stamp exchange. Furthermore, the measurement vector is 
\begin{equation}
\mathbf{z} = \begin{bmatrix} \tilde{\theta}_i^{k} \\ \tilde{\zeta}_i^k \end{bmatrix} = \begin{bmatrix}
1 & 0 \\ 0 & 1
\end{bmatrix}\begin{bmatrix}
\theta_i^k\\ \zeta_i^k
\end{bmatrix}  + \begin{bmatrix}
\kappa_{\theta} \\ \kappa_{\zeta}
\end{bmatrix}.
\end{equation}
The parameters $\kappa_{\theta}$ and $\kappa_{\zeta}$ represent the noise in the measurement of $\theta_i^k$ and $\zeta_i^k$, respectively.
The measured value $\tilde{\zeta}_i^k$ can be obtained using (\ref{eq:zeta}) while the measured offset is calculated by \cite{giorgi2011performance}
 \begin{equation}
\tilde{\theta}_i^{k}=\frac{1}{2}\left[\left(\cj{2}{k}+\cj{3}{k}\right)-\left(\ci{1}{k}+\ci{4}{k}\right)\right].
\end{equation} 
The prediction equations can then be rewritten as 
\begin{align}
& \mathbf{x}_{k|k-1} =  \mathbf{A}\mathbf{x}_{k-1},  \\
& \mathbf{P}_{k|k-1} = \mathbf{AP}_{k-1}\mathbf{A}^T,
\end{align}
where $\mathbf{P}_{k-1}$ and $\mathbf{P}_{k|k-1}$ denote a prior and a posterior prediction covariance matrix, respectively. Moreover, we can write the update equations as follows:
\begin{align}
& \mathbf{K}_{k} = \mathbf{P}_{k|k-1} \left[\mathbf{P}_{k|k-1} + \mathbf{R}\right]^{-1}, \\
& \mathbf{x}_{k} = \mathbf{x}_{k|k-1} + \mathbf{K}_{k}\left(\mathbf{z}- \mathbf{x}_{k|k-1}\right),\\
&\mathbf{P}_{k}  = \left(\mathbf{I} - \mathbf{K}_{k}\right) \mathbf{P}_{k|k-1},
\end{align}
where $\mathbf{K}_{k}$ is the Kalman gain and $\mathbf{R}$ denotes the measurement noise covariance matrix and can be given by \cite{giorgi2011performance}
\begin{equation}
\mathbf{R} = \sigma^2\begin{bmatrix}
1 & \frac{1}{\Delta T} \\ \frac{1}{\Delta T} & \frac{2}{(\Delta T)^2}
\end{bmatrix}.
\end{equation}
The parameter $\sigma^2$ is calculated as explained in section \ref{sssec:var}.
\subsection{Network-wide synchronization based on FG}\label{sec:fg}
\subsubsection{Introduction}
\begin{figure}
\includegraphics[width=1\linewidth]{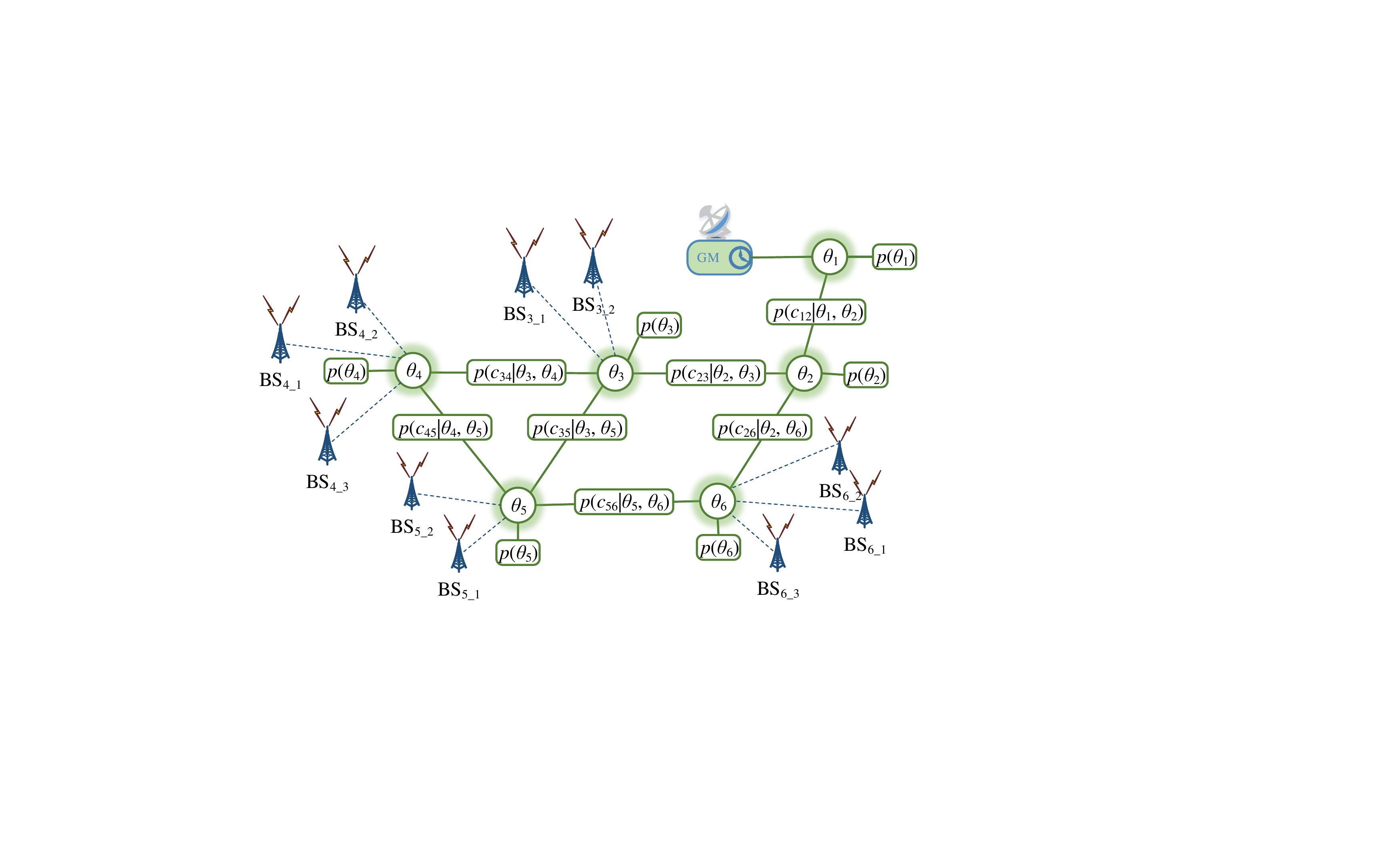}
\centering
\caption{Factor graph corresponding to an exemplary  communication network.}
\label{fig:fg}
\end{figure}

FGs are used to represent the factorization of probability distribution functions. As shown in Figure \ref{fig:fg} (the green graph), a FG comprises a number of nodes, each denoted by a variable and several factor nodes, each being a function of their neighboring variables. In particular, the factorization and graph structure in FGs can preserve the information about the form of the distribution while alleviating the computation load, e.g. that of marginal distribution through the sum-product algorithm \cite{kschischang2001factor}. 
Extending (\ref{eq:estpair}) to the whole network, the Bayesian posterior distribution of each node $\theta_i$ can be written as
\begin{align}
\begin{split}
&p(\theta_i|\{\Cij\}_{i=1:M, j\in ne(i)}) =\\& \int p(\theta_1,\cdots,\theta_M|\{\Cij\}_{i=1:M, j\in ne(i)}) d\theta_1\cdots d\theta_{i-1}d\theta_{i+1}\cdots d\theta_{M},\\
\end{split}
\label{eq:estnet}
\end{align} 
where
\begin{equation}
p(\theta_1,\cdots,\theta_M|\{\Cij\}_{i=1:M, j\in ne(i)}) \propto \prod p(\theta_i) \prod p(\Cij|\theta_i , \theta_j),
\label{eq:geq}
\end{equation}
and $ne(i)$ represents the set of neighboring nodes of node $i$. In general, the computation of the marginal in (\ref{eq:estnet}) is costly and of NP-hard complexity. However, there are several methods in the literature to efficiently represent the joint probability and compute the exact or approximate marginals. BP is an algorithm which can run on FG and relies on exchanging beliefs between neighboring nodes to compute the marginals. In the following, we briefly introduce the BP algorithm.
\begin{figure}
\includegraphics[width=0.95\linewidth]{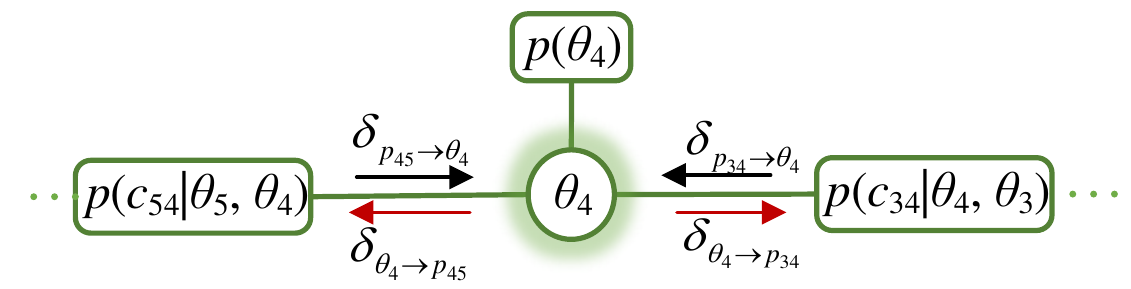}
\centering
\caption{Message passing in Belief Propagation.}
\label{fig:bpdet}
\end{figure}
\subsubsection{Belief Propagation}
Figure \ref{fig:bpdet} depicts the principles of message passing in the BP algorithm for the exemplary node $\theta_4$. For the sake of simplicity, we denote the factor $p(\Cij|\theta_i, \theta_j)$ by $p_{ij}.$ The message from a variable vertex $\theta_i$ to factor vertex $\pij$ is given by \cite{barberBRML2012}
\begin{equation}
\mij{\theta_i}{\pij}(\theta_i) = p(\theta_i)\prod_{p_{ik}\in\{ne(\theta_i)\setminus \pij\}} \mij{p_{ik}}{\theta_i}(\theta_i),
\label{eq:bpmsg1}
\end{equation}
where $\mij{p_{ik}}{\theta_i}(\theta_i)$ is the message from a factor vertex $\pij$ to the variable vertex $\theta_i,$ and is given by
\begin{equation}
\mij{\pij}{\theta_i}(\theta_i) = \int\ p(\Cij|\theta_i, \theta_j) \mij{\theta_j}{\pij}(\theta_j)d\theta_j.
\label{eq:bpmsg2}
\end{equation}
It is straightforward to see that
\begin{equation}
b(\theta_i) \propto  p(\theta_i)\prod_{p_{ik}\in ne(\theta_i)} \mij{p_{ik}}{\theta_i}(\theta_i),
\label{eq:bel}
\end{equation} 
where $b(\theta_i)$ denotes the marginal belief of variable node $\theta_i$. 
The BP procedure can be summarized as  
\begin{enumerate}
\item The message $\mij{\theta_i}{\pij}(\theta_i)$ is transmitted from $\theta_i$ to the neighboring factor node $\pij$ (it is initialized non-informatively in the first iteration),
\item The factor node $\pij$ compute the message $\mij{\pij}{\theta_i}(\theta_i)$ based on its incoming messages and send the calculated message to the neighboring node $\theta_i,$ 
\item Each node $i$ updates its belief $b(\theta_i)$ based on the received messages from the neighboring factor nodes. 
\end{enumerate}
We note that in practice there are neither factors nor variable nodes meaning that (\ref{eq:bpmsg1}) and (\ref{eq:bpmsg2}) are calculated locally at each node and only $\mij{p_{ik}}{\theta_i}(\theta_i)$ is sent to the neighboring node $i$.
\begin{algorithm}[t!]
\begin{algorithmic}[1]
\STATE Determine the suitable algorithm for each part of the network (BP-nodes or KF-nodes). \label{detalg}
\STATE Start the time-stamping exchange and correspondingly the KF algorithm at KF-nodes. \label{klex}
\STATE Start the time-stamp exchange between adjacent BP-nodes and Calculate (\ref{eq:condprob}) for each pair \label{msgex}
\FOR {$l = 1, 2, \cdots, L$}  \label{for}
\STATE Compute the messages using (\ref{eq:bpmsg1}) and (\ref{eq:bpmsg2}) for each BP-node and transmit them to its neighboring nodes \label{calmsg}
\STATE Compute the offset estimation at each BP-node using (\ref{eq:bel}) and update their belief \label{caloff}
\IF{$\hat{\theta}_i^{(l)}-\hat{\theta}_i^{(l-1)}\leq\epsilon\ \forall i$} \label{conv} 
\STATE Go to step \ref{msgex}
\ENDIF \label{convend} 
\ENDFOR \label{endfor}
\end{algorithmic}
\caption{Network synchronization algorithm}
\label{alg:hybalg}
\end{algorithm}
\subsection{Hybrid BP-KF}\label{ssec:hyb}
Given Sections \ref{sec:kf} and \ref{sec:fg}, one can decide on the suitable algorithm for each sync domain in network. That is, the nodes backhauling the BSs need to be precisely synchronized using BP whereas the nodes at the edge of the network (BSs) can be synchronized using KF where mostly global sync is of less importance compared to local sync precision.

Algorithm \ref{alg:hybalg} describes the steps of the hybrid synchronization approach. Firstly, in step \ref{detalg}, we decide on the network sections where BP and KF are to be applied (they are labeled as BP-nodes and KF-nodes, respectively). Later, in step \ref{klex}, the time-stamp exchange mechanism shown in Figure \ref{fig:stamp} and, correspondingly, the KF algorithm is initiated at the KF-nodes. In step \ref{msgex}, the time-stamp exchange is initiated among the BP-nodes, thereby obtaining the required time-stamps to calculate the conditional probability in (\ref{eq:condprob}). The BP iterations begin at step \ref{for} and continue until convergence or when the maximum number of iterations $L$ is reached. In step \ref{calmsg}, each BP-node calculates its outgoing messages and sends them to their corresponding nodes. Each node's belief is then computed in step \ref{caloff} using (\ref{eq:bel}). Steps \ref{conv}-\ref{convend} are responsible to check the convergence by comparing the difference of clock offset estimations in iterations $(l)$ and $(l-1)$ with a predefined small value $\epsilon$. It is noteworthy that the step \ref{klex} and steps \ref{msgex}-\ref{endfor} can continuously run in parallel.
\section{Simulation Results}
We consider the network in Figure \ref{fig:fg} as an exemplary scenario, where a number of BSs are backhauled by a mesh network.  We conduct two sets of simulations: a) synchronizing the whole network based only on FG and, correspondingly, BP algorithm (the BSs in Figure \ref{fig:fg} are assumed to be variable nodes as well and connected to the mesh network via factors), and b) we perform synchronization in a hybrid manner where the mesh backhauling network is synchronized based on FG while the BSs at the edge of the network are being synchronized using KF. We then compute Root Mean Square Error (RMSE) of offset estimation as a measure to evaluate the performance in each scenario. For the sake of simplicity and without loss of generality we consider only the nodes $\theta_4$ and $\theta_6$ and their corresponding BSs. Moreover, the simulation parameters are set as in Table \ref{tab:sim} and the Python package in \cite{pyfac} is employed to perform the message passing algorithm.
\begin{table}[t!]
\centering
\caption{Simulation parameters}
\begin{tabular}{l|c}
Number of independent simulations &	10000 \\ \hline
Initial random delays &	[-50, 50] ns \\ \hline
Number of time-stamp exchange $K$&	10 \\ \hline
Standard deviation of $T_k$ and $R_k$ &	4 ns \\ \hline
Random delay between each pair of nodes &	$\left[ 200, 300 \right]$ ns \\ \hline
Initial pdf of the offset for each node & $\mathcal{N}(0,+\infty)$\\\hline
Initial pdf of the offset of MN & $\mathcal{N}(0,0)$
\end{tabular}
\label{tab:sim}
\end{table}
\begin{figure}[t!]
\includegraphics[width=0.85\linewidth]{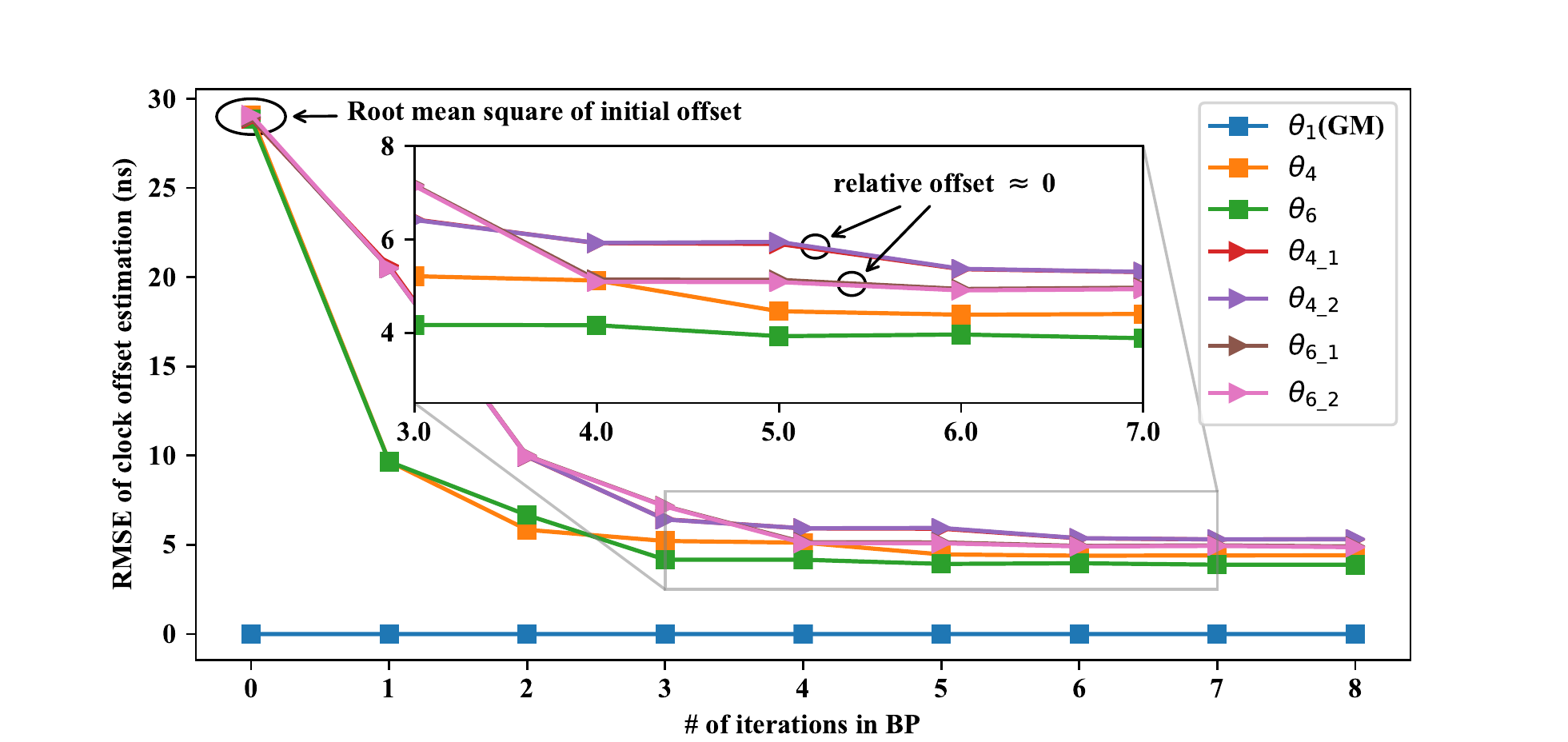}
\centering
\caption{BP applied on the whole network (scenario a).}
\label{fig:bpwhole}
\end{figure}
\begin{figure}[t!]
\includegraphics[width=0.84\linewidth]{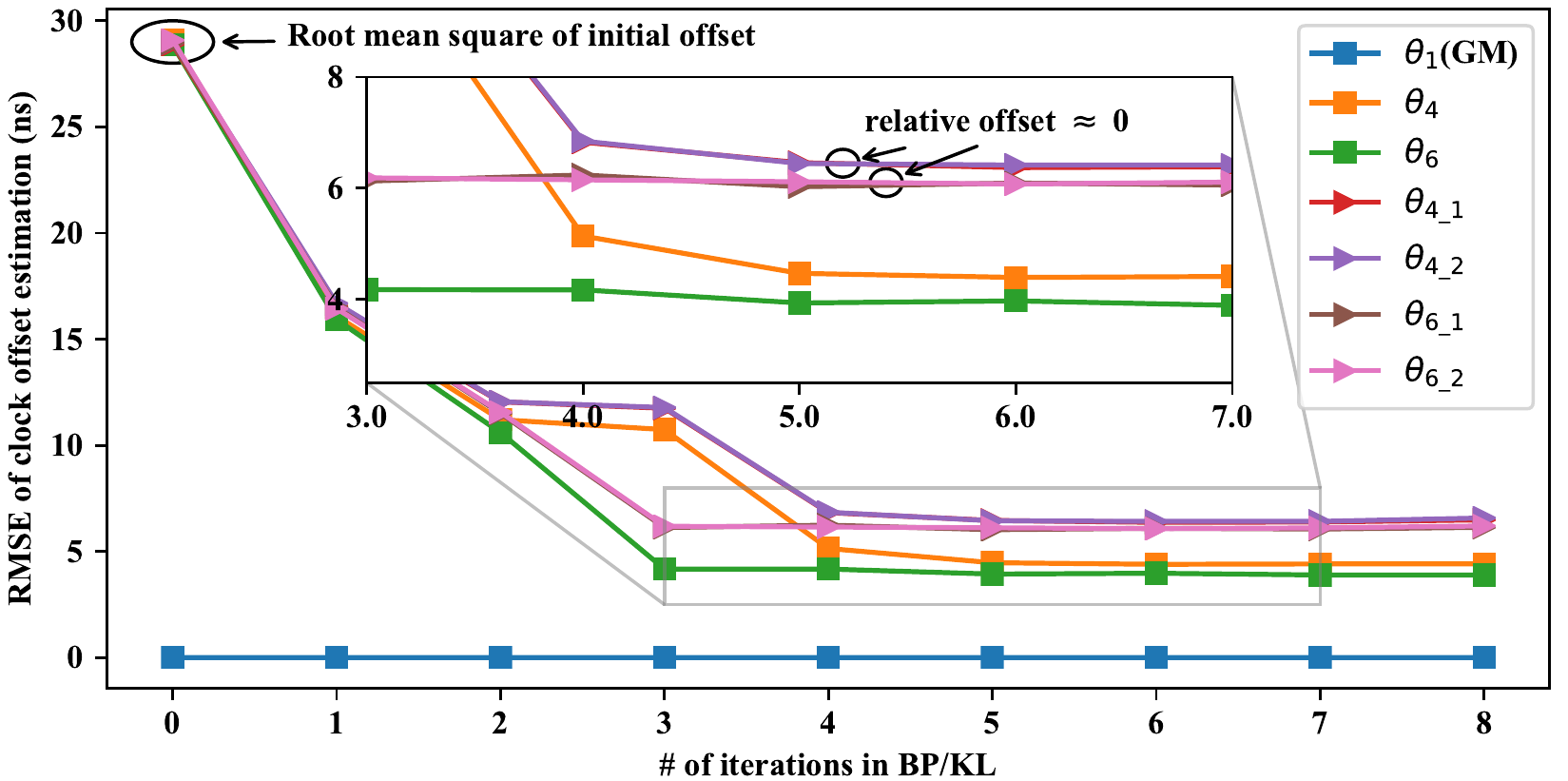}
\centering
\caption{BP applied only to the backhauling mesh network (scenario b).}
\label{fig:bpmesh}
\end{figure}

Figure \ref{fig:bpwhole} represents RMSE of offset estimation for scenario (a). As can be seen, the BP converges after $4$ iterations achieving a synchronization accuracy of $4-6$ ns with respect to the Grand Master (GM) node. The convergence is guaranteed for networks with at least one GM \cite{leng2011distributed}. However, when a network contains loops, the value to which BP converges, is considered to be approximate \cite{barberBRML2012}. 
In fact, the results in this simulation setup reveal the potential performance of BP for time synchronization in communication networks. However, the nodes, and particularly the BSs, must wait at least $4$ iterations (or more if a network is more complex) to be completely synchronized. This can be problematic in a number of applications, e.g., localization, where continuous time alignment is essential. Therefore, it appears necessary for the BSs to synchronize themselves more frequently to be able to deliver certain services.    

Figure \ref{fig:bpmesh} shows the RMSE of offset estimation for scenario (b). As can be observed, the performance slightly deteriorates (around $2$ ns) compared to scenario (a). However, we note that the iterations of KF are significantly faster than that of BP. In fact, BP begins only when the nodes have already conducted several rounds of time-stamp exchange (in order to obtain the conditional probabilities) and, even then, it still needs 4 iterations to perform synchronization. In contrast, KF updates the estimation after each round of time-stamp exchange, thereby maintaining the relative clock offsets low. In other words, since the KF is faster and runs independently (does not need any information from the other network sections as BP does), it is able to conduct more iterations, thereby continuously fulfilling the local requirement of relative time error.

In summary, the simulation results indicate that BP can be of great potential for high precision network synchronization. Nevertheless, despite the excellent performance, high number of message passing iterations can cause trouble by prolonging the sync period. In particular, the time needed for the nodes to exchange time-stamps and pass messages can lead to deterioration in accuracy of synchronization. As a solution, the hybrid approach explained in section \ref{ssec:hyb} can be adopted to alleviate the above-mentioned problem. That is, applying BP only on the critical parts of the network (e.g., the backhauling part which is responsible for distributing the clock to the edges) to achieve as high a precision as possible in global level. Moreover, faster algorithms, e.g. KF, can be readily employed on the edges of the network where precise, fast, and frequent local synchronization is required for numerous applications

\section{Conclusion and Future Work}
We presented two algorithms to synchronize the nodes in communication networks, each extensively discussed and shown to have benefits and drawbacks. One is based on Factor Graphs and able to achieve extremely accurate synchronization with higher complexity (high number of time-stamp exchanges and message passing iterations), while the other can deliver strong performance in tree structure networks. Further on, we combined the two approaches to maintain synchronization accuracy on a global level while performing frequent precise synchronization at local level. Simulation results show that the proposed hybrid network can achieve high precision and frequent synchronization at the cost of a slight deterioration in performance.

We only dealt with clock offset estimation, however skew compensation cannot be ignored while designing a sustainable synchronization algorithm. The future works aim at incorporating skew synchronization into the proposed algorithm to further enhance the performance.

%
\bibliography{synch_paper_1}
\bibliographystyle{IEEEtran}

\end{document}

%% file: Macros.tex
\newcommand*{\argmax}{\ensuremath{\mathop{\mathrm{arg\,max}}}}
\newcommand{\Cij}{\mathbf{c}_{ij}}
\newcommand{\Cijk}[1]{c_{ij}^{#1}}
\newcommand{\ci}[2]{c_{j,#1}^{#2}}

\newcommand{\alij}{\alpha_{ij}}
\newcommand{\cj}[2]{c_{i,#1}^{#2}}
\newcommand{\mij}[2]{\delta_{#1\rightarrow #2}}
\newcommand{\pij}{p_{ij}}

\newcommand{\cntn}[2]{p(\Cijk{#1}|\theta_i^{#2})}

\newcommand{\tnt}[2]{p(\theta_i^{#1}|\theta_i^{#2})}

\newcommand{\jthnc}[2]{p(\theta_i^{#1}, \cdots, \theta_i^{#2}|\Cij)}

\newcommand{\jcthn}[2]{p(\Cij|\theta_i^{#1}, \cdots, \theta_i^{#2})}
\newcommand{\tta}[1]{\theta_i^{#1}}
\newcommand{\ttan}[2]{p(\theta_i^{#1}, \cdots, \theta_i^{#2})}

\newcommand{\cij}{\mathbf{c}_{i\rightarrow j}}
\newcommand{\cji}{\mathbf{c}_{j\rightarrow i}}